\documentclass[%
preprint,
frontmatterverbose, 
 amsmath,amssymb,
 aps,
prb,
floatfix,
]{revtex4-1}

\usepackage{color}
\usepackage{graphicx}
\usepackage{dcolumn}
\usepackage{bm}
\usepackage{hyperref}
\usepackage[mathlines]{lineno}

\begin{document}

\preprint{APS/123-QED}

\title{Spin splitting with persistent spin textures induced by the line defect in 1T-phase of monolayer transition metal dichalcogenides}

\author{Moh. Adhib Ulil Absor}
\email{adib@ugm.ac.id} 
\affiliation{Department of Physics, Faculty of Mathematics and Natural Sciences, Universitas Gadjah Mada, Sekip Utara BLS 21 Yogyakarta 55186, Indonesia.}%

\author{Iman Santosa}
\affiliation{Department of Physics, Faculty of Mathematics and Natural Sciences, Universitas Gadjah Mada, Sekip Utara BLS 21 Yogyakarta 55186, Indonesia.}%

\author{Naoya Yamaguchi}%
\affiliation{Nanomaterial Reserach Institute (NANOMARI), Kanazawa University, 920-1192 Kanazawa, Japan.}%

\author{Fumiyuki Ishii}%
\affiliation{Nanomaterial Reserach Institute (NANOMARI), Kanazawa University, 920-1192 Kanazawa, Japan.}%

\date{\today}

\begin{abstract}

The spin splitting driven by spin-orbit coupling in monolayer (ML) transition metal dichalcogenides (TMDCs) family has been widely studied only for the 1H-phase structure, while it is not profound for the 1T-phase structure due to the centrosymmetric of the crystal. Based on first-principles calculations, we show that significant spin splitting can be induced in the ML 1T-TMDCs by introducing the line defect. Taking the ML PtSe$_{2}$ as a representative example, we considered the most stable form of the line defects, namely Se-vacancy line defect (Se-VLD). We find that large spin splitting is observed in the defect states of the Se-VLD, exhibiting a highly unidirectional spin configuration in the momentum space. This peculiar spin configuration may yield the so-called persistent spin textures (PST), a specific spin structure resulting in protection against spin-decoherence and supporting an extraordinarily long spin lifetime. Moreover, by using $\vec{k}\cdot\vec{p}$ perturbation theory supplemented with symmetry analysis, we clarified that the emerging of the spin splitting maintaining the PST in the defect states is originated from the inversion symmetry breaking together with one-dimensional nature of the Se-VLD engineered ML PtSe$_{2}$. Our findings pave a possible way to induce the significant spin splitting in the ML 1T-TMDCs, which could be highly important for designing spintronic devices.

\end{abstract} 

\maketitle

\section{INTRODUCTION}

Since the experimental isolation of graphene in 2004\cite{Novoselov}, significant research efforts have been devoted to the investigation of two-dimensional (2D) materials with atomically-thin crystals\cite{Tan}. Here, growing research attention has been focused on the monolayer (ML) transition metal dichalcogenides (TMDCs) family due to a high possibility to be used in future nanoelectronic devices\cite {Manzeli}. Most of the ML TMDCs families have graphene-like hexagonal crystal structures where transition metal atoms ($M$) are sandwiched between layers of chalcogen atoms ($X$) with $MX_{2}$ stoichiometry. However, due to the local coordination of the transition metal atoms, they admit two different stable forms in the ground state, namely a 1H-phase structure having trigonal prismatic symmetry, and a 1T-phase structure that consists of distorted octahedral symmetry \cite {Cudazzo}. The different coordination environments in the ML TMDCs lead to distinct crystal field splitting of the $d$-like bands. However, depending on the transition metal atom species, the ML TMDCs display metallic, semiconducting, or insulator behavior\cite{Wilson}. Therefore, various physical properties such as tunability of bandgap \cite {Absor, Ramasubramaniam}, high carrier mobility\cite{YWang, XZhang}, and superior surface reactivity \cite{Chia} are established, evidencing that the ML TMDCs is an ideal platform for next-generation technologies.
 
Of special interest is the promising application of the ML TMDCs for spintronics devices due to the strong spin-orbit coupling (SOC), which is particularly noticeable in the ML 1H-TMDCs such as ML (Mo/W)$X_{2}$ ($X$ = S, Se)\cite {Zhu, Liu_Bin, Absor}. Here, the lack of the crystal inversion symmetry together with the strong SOC in the 5$d$ orbitals of transition metal atoms leads to the large spin splitting in the electronic band structures. This effect is conspicuously apparent in the valence band maximum exhibiting the spin splittings ranging between 150 meV (ML MoS$_{2}$) up to 400 meV (WSe$_{2}$)\cite{Zhu, Liu_Bin, Absor}. Due to the well separated valleys at the $K$ and $K'$ points in the hexagonal Brillouin zone, this splitting gives rise to the so-called spin-valley coupling\cite{Xiao_b}, which is responsible for the appearance of valley-contrasting effects such as spin Hall effect \cite{Cazalilla}, spin-dependent selection rule for optical transitions \cite{Chu}, and magneto-electric effect in the ML TMDCs \cite {Gong}. Furthermore, an electrically controllable spin splitting and spin polarization in the ML 1H-TMDCs has been reported \cite{KGong}, making them suitable for spin-field effect transistor.         

Compared to the well-studied ML 1H-TMDCs, the effect of the SOC in the ML 1T-TMDCs is equally interesting. Especially the ML PtSe$_{2}$ has attracted much scientific attention since it has been successfully synthesized by a direct selenization at the Pt(111) substrate\cite{YWang, Yao}. Moreover, this material has been predicted to exhibit the largest electron mobility among the widely studied ML TMDCs \cite{YWang, XZhang}. Recently, Yao $et$. $al$.\cite{Yao}, by using spin- and angle-resolved photoemission spectroscopy (spin-ARPES), reported spin-layer locking phenomena in the ML PtSe$_{2}$; that is, the spin-polarized states are degenerated in energy but spatially locked into two sublayers forming an inversion partner. Similar phenomena has also been theoretically predicted on other ML 1T-TMDCs such as ML (Zr/Hf)$X_{2}$ ($X$ = S, Se)\cite{CaiC}. This phenomena, which is a manifestation of the global centrosymmetric of the crystal and the local dipole-induced Rashba SOC effect, may provide a disadvantage for spintronics applications. Since the ML 1T-TMDCs possesses superior transport properties due to the high electron mobility \cite{YWang, XZhang}, lifting the spin degeneracy in the ML 1T-TMDCs could be the important key for their realization in the spintronics devices. Therefore, finding a feasible method to induce the significant spin splitting in the ML 1T-TMDCs is highly desirable.           

In this paper, by using density-functional theory (DFT) calculations, we show that the significant spin splitting can be induced in the ML 1T-TMDCs by introducing the line defect. By using the ML PtSe$_{2}$ as a representative example, we investigate the most stable form of the line defects, namely Se-vacancy line defect (Se-VLD). We find that a sizable spin splitting is observed in the defect states of the Se-VLD, exhibiting a highly unidirectional spin configuration in the momentum space. This peculiar spin configuration gives rise to the so-called persistent spin textures (PST)\cite{Schliemann, Bernevig}, a specific spin structure that protects the spin from decoherence and induces an extremely long spin lifetime\cite {Dyakonov, Altmann}. Moreover, by using $\vec{k}\cdot\vec{p}$ perturbation theory supplemented with symmetry analysis, we clarified that the emerging of the spin splitting maintaining the PST in the defect states is originated from the inversion symmetry breaking and one-dimensional (1D) nature of the Se-VLD engineered ML PtSe$_{2}$. Finally, a possible application of the present system for spintronics will be discussed.

\section{Computational Details}

We performed first-principles electronic structure calculations based on the DFT within the generalized gradient approximation (GGA) \cite {Perdew} implemented in the OpenMX code \cite{Openmx}. Here, we adopted norm-conserving pseudopotentials \cite {Troullier} with an energy cutoff of 350 Ry for charge density. The wave functions are expanded by the linear combination of multiple pseudoatomic orbitals (LCPAOs) generated using a confinement scheme \cite{Ozaki, Ozakikino}. The orbitals are specified by Pt7.0-$s^{2}p^{2}d^{2}$ and Se9.0-$s^{2}p^{2}d^{1}$, which means that the cutoff radii are 7.0 and 9.0 Bohr for the Pt and Se atoms, respectively, in the confinement scheme \cite{Ozaki, Ozakikino}. For the Pt atom, two primitive orbitals expand the $s$, $p$, and $d$ orbitals, while, for the Se atom, two primitive orbitals expand the $s$ and $p$ orbitals, and one primitive orbital expands $d$ orbital. The SOC was included in the DFT calculations by using $j$-dependent pseudopotentials \citep{Theurich}. The spin textures in the momentum space were calculated using the spin density matrix of the spinor wave functions obtained from the DFT calculations as we applied recently on various 2D materials \cite{Absor3, Absor4}.  

\begin{figure}
	\centering
		\includegraphics[width=0.85\textwidth]{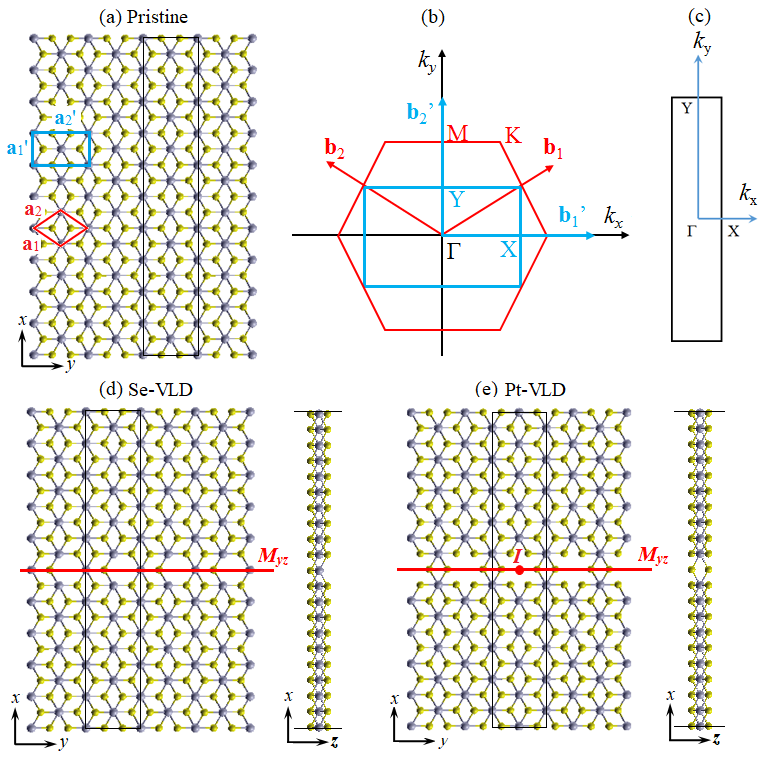}
	\caption{(a) Top view of the crystal structures of the pristine ML PtSe$_{2}$ where red, blue, and black lines represent the primitive hexagonal cell, minimum rectangular cell, and rectangular supercell, respectively, corresponding to the folding of the FBZ between primitive hexagonal and minimum rectangular cells (b). (c) The FBZ for the pristine supercell is also shown. Top and side views of the crystal for: (d) Se-vacancy line defect (Se-VLD) and (e) Pt-vacancy line defect (Pt-VLD). Both the line defects are oriented along the armchair direction ($y$-direction), while they are isolated in the direction perpendicular to the line defect ($x$-direction) as indicated by black lines of the rectangular lattice. Here, all of the symmetry operations including the inversion symmetry ($I$) and mirror symmetry $M_{yz}$ are indicated by the red point and lines.}
	\label{figure:Figure1}
\end{figure}

To model the VLD in the ML 1T-TMDCs, we considered the ML PtSe$_{2}$ as a representative example. Here, we constructed a supercell of the pristine ML PtSe$_{2}$ from the minimum rectangular cell [Fig. 1(a)], where the optimized lattice parameters are obtained from the primitive hexagonal cell. As a consequence, the folding cell from the hexagonal to rectangular cells in the FBZ is expected as shown in Fig. 1(b)-(c). Here, we used the axes system where the ML is chosen to sit on the $x-y$ plane, where the $x$ ($y$) axis is taken to be parallel to the zigzag (armchair) direction. We considered two different configurations of the VLD, namely the Se-VLD and Pt-VLD, where their relaxed structures are displayed in Figs. 1(d) and 1(e), respectively. To model these VLDs, we extend the supercell size of the ML PtSe$_{2}$ by ten times in the $x$-direction, which is perpendicular to the direction of the extended vacancy line along the $y$-direction to eliminate interaction between the periodic image of the line defect [see Fig. 1(d)-(e)].

In our DFT calculations, we used a periodic slab where a sufficiently large vacuum layer (20 \AA) is used to avoid interaction between adjacent layers. The $3\times12\times1$ k-point mesh was used, and the geometries were fully relaxed until the force acting on each atom was less than 1 meV/\AA. To confirm energetic stability of the VLD, we calculate vacancy formation energy ($E_{f}$) through the following relation \cite {Freysoldt}:
\begin{equation}
\label{1}
E_{f}=E_{\texttt{VLD}}-E_{\texttt{Pristine}}+\sum_{i}n_{i}\mu_{i}.
\end{equation}
In Eq. (1), $E_{\texttt{VLD}}$ is the total energy of the VLD, $E_{\texttt{Pristine}}$ is the total energy of the pristine system, $n_{i}$ is the number of atom being removed from the pristine system, and $\mu_{i}$ is the chemical potential of the removed atoms corresponding to the chemical environment surrounding the system. Here, $\mu_{i}$ obtains the following requirements:
\begin{equation}
\label{2}
E_{PtSe_{2}}-2E_{Se}\leq \mu_{Pt}\leq E_{Pt},
\end{equation}
\begin{equation}
\label{3}
\frac{1}{2}(E_{PtSe_{2}}-E_{Pt})\leq \mu_{Se}\leq E_{Se}.
\end{equation}
Under Se-rich condition, $\mu_{Se}$ is the energy of the Se atom in the bulk phase (hexagonal Se, $\mu_{Se}=\frac{1}{3}E_{Se-hex}$) which corresponds to the lower limit on Pt, $\mu_{Pt}=E_{PtSe_{2}}-2E_{Se}$, where $E_{PtSe_{2}}$ is the total energy of the ML PtSe$_{2}$ in the primitive unit cell. On the other hand, in the case of the Pt-rich condition, $\mu_{Pt}$ is associated with the energy of the Pt atom in the bulk phase (fcc Pt, $\mu_{Pt}=\frac{1}{4}E_{Pt-fcc}$) corresponding to the lower limit on Se, $\mu_{Se}=\frac{1}{2}(E_{PtSe_{2}}-E_{Pt})$.

\section{RESULT AND DISCUSSION}

First, we briefly discuss the structural symmetry and electronic properties of the pristine ML PtSe$_{2}$. 
The ML PtSe$_{2}$ crystallizes in a centrosymmetric crystal associated with the ML 1T-TMDCs with the $P\bar{3}mI$ space group for the global structure. Here, one Pt atom (or Se atom) is located on top of another Pt atom (or Se atom) forming octahedral coordination, while it shows trigonal structure when projected to the (001) plane [Fig. 1(a)]. As a result, a polar group $C_{3v}$ and a centrosymmetric group $D_{3d}$ are identified for the Se and Pt sites, respectively. We find that the calculated lattice constant of the pristine ML PtSe$_{2}$ in the primitive unit cell is 3.75 \AA, which is in good agreement with previous theoretical (3.75 \AA \cite {Zulfiqar, WZhang}) and experimental (3.73 \AA \cite {YWang}) results.

\begin{figure*}
	\centering
		\includegraphics[width=0.8\textwidth]{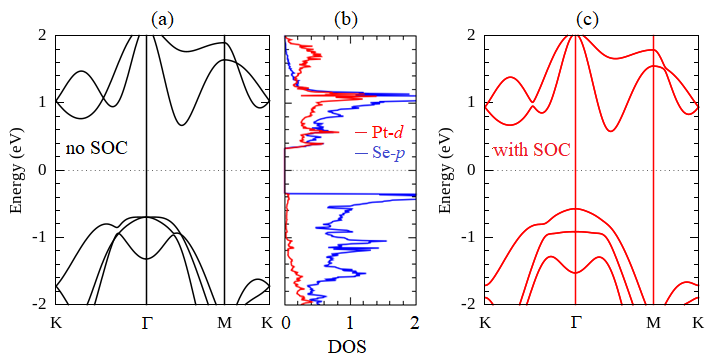}
	\caption{(a) Electronic band structure of the pristine ML PtSe$_{2}$ calculated without the spin-orbit coupling (SOC) corresponding to (b) density of states projected to the atomic orbitals. (c) the electronic band structures calculated with including the SOC.}
	\label{figure:Figure2}
\end{figure*}

Figure 2(a) shows the calculated result of the electronic band structures of the ($1 \times 1$) primitive unit cell of the pristine ML PtSe$_{2}$. Consistent with previous results \cite{WZhang, YWang}, we find that the pristine ML PtSe$_{2}$ is an indirect semiconductor with the bandgap of 1.38 eV, where the valence band maximum (VBM) is located at $\Gamma$ point, while the conduction band minimum (CBM) is located at the $\vec{k}$ along the $\Gamma-M$ line. Our calculated results of the density of states (DOS) projected to the atomic orbitals confirmed that the VBM is predominantly contributed from the Se-$p$ orbitals, while the CBM is mainly originated from the Pt-$d$ orbitals [Fig. 2(b)]. Turning the SOC, lifting spin degeneracy of the electronic band structure is expected, which is driven by the lack of the inversion symmetry \cite {Rashba, Dresselhauss}. However, in the ML PtSe$_{2}$, we find that all the bands are doubly degenerated, which is protected by the centrosymmetric of the crystal [Fig. 2(c)]. Evidently, the spin degeneracy observed in the electronic band structures of the pristine ML PtSe$_{2}$ is consistent with the recent experimental results reported by Yao $et$. $al$., by using  spin-ARPES\cite{Yao}.   

\begin{table}[ht!]
\caption{The calculated vacancy formation energy $E_{f}$ (in eV) of the Se-VLD and Pt-VLD in the ML PtSe$_{2}$ corresponding to the Pt-Se bond length $d_{\textbf{Pt-Se}}$ (in \AA) around the vacancy cite, compared with that of the single vacancy defect (Se-SVD and Pt-SVD) and the pristine systems. Several theoretical data of the SVD systems from the previous reports are presented for a comparison.} 
\centering 
\begin{tabular}{c c c c} 
\hline\hline 
Defective systems & $E_{f}$ (eV)    & $d_{\textbf{Pt-Se}}$ (\AA) & Ref. \\ 
\hline 
Pristine &   & 2.548 &  This work \\
Se-VLD & 1.92 (Se-rich); 1.46 (Pt-rich)   & 2.553 & This work \\ 
Pt-VLD & 4.85 (Se-rich); 4.11 (Pt-rich) & 2.576 & This work \\
Se-SVD & 1.78 (Se-rich); 1.35 (Pt-rich)  & 2.509 & This work \\
       & 1.84 (Se-rich); 1.27 (Pt-rich)  & 2.509 & Ref. \cite{AbsorA} \\
       & 1.87 (Se-rich); 1.24 (Pt-rich)  &   & Ref. \cite{WZhang} \\
Pt-SVD & 4.35 (Se-rich); 3.11 (Pt-rich)   & 2.508 & This work \\         
       & 4.28 (Se-rich); 3.06 (Pt-rich)  &  2.508  &   Ref. \cite{AbsorA} \\
       & 4.19 (Se-rich); 3.00 (Pt-rich)  &  & Ref. \cite{WZhang} \\
\hline\hline 
\end{tabular}
\label{table:Table 1} 
\end{table}

When the VLD is introduced, the position of the atoms around the VLD site significantly changes from that of the pristine atomic position due to the relaxation. To examine the optimized structures of the VLD, we show the calculated results of the Pt-Se bond length $d_{\textbf{Pt-Se}}$ around the VLD site in Table 1. In the case of the Se-VLD, removing one Se atom from the supercell breaks the inversion symmetry ($I$) of the ML PtSe$_{2}$ [Fig. 1(d)]. Consequently, three Pt atoms around the VLD site are relaxed moving close to each other, implying that $d_{\textbf{Pt-Se}}$ at each hexagonal side around the VLD site has the same value of about 2.553 \AA. As a result, a mirror symmetry plane ($M_{yz}$) exist along the extended vacancy line [Fig. 1(d)]. In contrast to the Se-VLD case, the Pt-VLD retains both the inversion symmetry $I$ and mirror symmetry $M_{yz}$ in the supercell [Fig. 1(e)]. However, the geometry of the Pt-VLD undergoes significant distortion from the pristine crystal, resulting in that $d_{\textbf{Pt-Se}}$ of the Pt-VLD (2.576 \AA) is slightly larger than that of the Se-VLD (2.553 \AA) and the pristine (2.548 \AA) systems.

\begin{figure*} 
	\centering
		\includegraphics[width=1.0\textwidth]{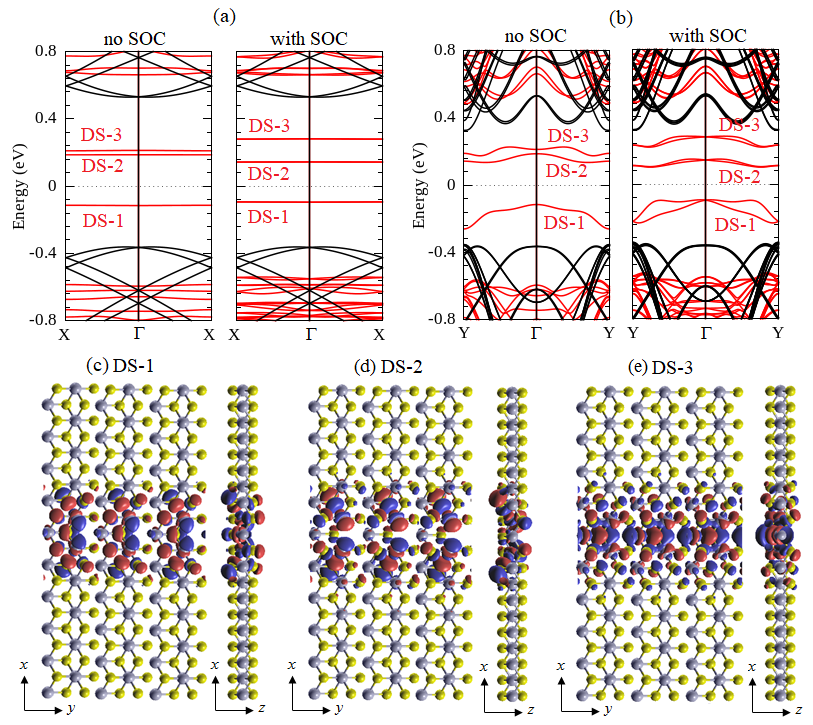}
	\caption{Band structures calculated along (a) $X-\Gamma-X$  and (b) $Y-\Gamma-Y$ directions for the Se-VLD (red lines) compared with those of the pristine supercell systems (black lines) without (left) and with (right) including the SOC. Here, DS-1 indicates an occupied defect state, while DS-2 and DS-3 represent the two unoccupied defect states. Isosurface map of the wave function calculated at the $\Gamma$ point for (c) the DS-1, (d) the DS-2, and (e) the DS-3 defect states, where the isovalue of the wave function is 0.02 e/\AA$^{3}$. }
	\label{figure:Figure3}
\end{figure*}

To assess the stability of the proposed VLDs, we calculate their formation energy $E_{f}$. As shown in Table 1, we find that the calculated $E_{f}$ of the Se-VLD is much smaller than that of the Pt-VLD under the Se-rich and Pt-rich conditions, indicating that the Se-VLD is easily formed in the ML PtSe$_{2}$. In contrast, the formation of the Pt-VLD is highly unfavorable due to the required energy. Since the Pt atom is covalently bonded to the six neighboring Se atoms, removing the Pt atom in the supercell is stabilized by increasing the $E_{f}$. For a comparison, we also calculate $E_{f}$ of a Se-single vacancy defect (Se-SVD) and a Pt-single vacancy defect (Pt-SVD) by using the same supercell model. We find that the calculated $E_{f}$ of the Se-VLD is comparable to that of the Se-SVD, but is much smaller than that of the Pt-SVD [see Table 1], suggesting that the formation of the Se-VLD is energetically accessible. The found stability of the Se-VLD is consistent with the previous report that the chalcogen vacancies and their alignment into the extended line defects can be easily formed in the ML TMDCs \cite{Komsa, Noh, Li, AbsorA, WZhang}. In fact, aggregation of the chalcogen vacancy and their alignment into extended line defect has been experimentally reported on the ML MoS$_{2}$ by using electron irradiation technique\cite{Komsa1, ChenQ, WangS}, indicating that the Se-VLD engineered ML PtSe$_{2}$ is experimentally feasible. Since the Se-VLD has the lowest formation energy in the VLD systems, in the following discussion, we will concentrate only on the electronic properties of the Se-VLD engineered ML PtSe$_{2}$.

Figures 3(a) and 3(b) show the band structures of the Se-VLD along the $\Gamma-X-\Gamma$ and $\Gamma-Y-\Gamma$ lines in the FBZ, respectively, corresponding to the electronic wavefunction [Figs. 3(c)-3(e)] at the $\Gamma$ point around the Fermi level. Without the SOC, we identify three defect levels inside the bandgap, which is characterized by a single occupied defect state (DS-1) and two unoccupied defect states (DS-2 and DS-3). Our spin-polarized calculations on the Se-VLD system confirmed that there is no lift of the spin degeneracy in the defect states, indicating that the Se-VLD engineered ML PtSe$_{2}$ remains non-magnetic as the defect-free one. The observed defect states, together with the non-magnetic character of the Se-VLD engineered ML PtSe$_{2}$, are consistent with previous results of a single chalcogen vacancy in various ML TMCDs\cite{Li, AbsorA, Noh}. Importantly, we observe a dispersive character of the bands in the defect states, which is visible in the $\Gamma-Y$ direction [Fig. 3(b)]. However, the bands are dispersionless in the defect states along the $\Gamma-X$ direction [Fig. 3(a)]. The dispersionless defect states along the $\Gamma-X$ direction is because the Se-VLD is completely isolated far from each other in the direction perpendicular to the vacancy line [see Fig. 1(d)]. On the other hand, the interaction between the neighboring vacancies along the extended vacancy line induces delocalized wave function forming a quasi-1D confined state along the vacancy line direction [Figs. 3(c)-3(e)]. Such peculiar wave function is responsible for inducing the dispersive defect states along the $\Gamma-Y$ direction, which is expected to enhance carrier mobility\cite{Fishchuk}, and thus plays an important role in the transport-based electronic devices.

\begin{figure*}
	\centering
		\includegraphics[width=1.0\textwidth]{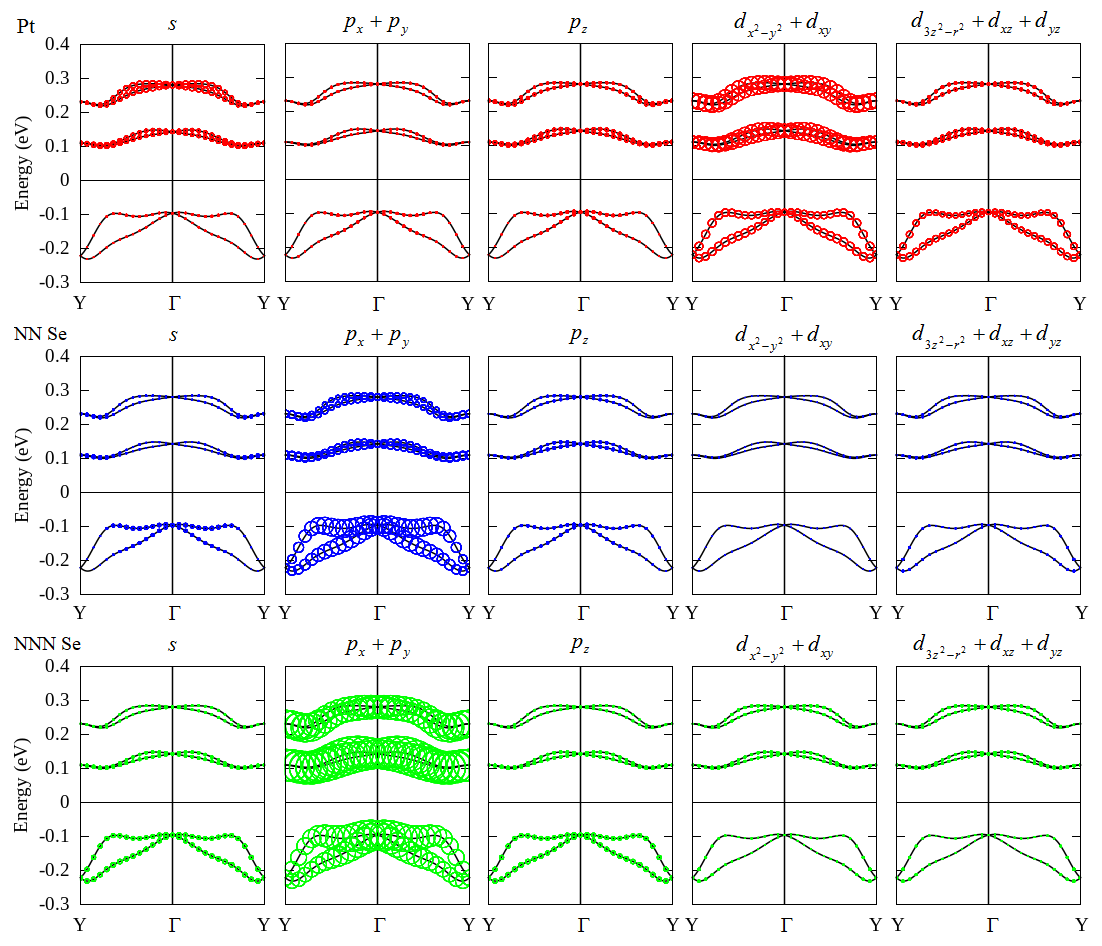}
	\caption{Orbital-resolved of electronic band structures calculated at the defect states projected to the Pt (red circles), the nearest neighbor (NN) Se (blue circles), and the next-nearest neighbor (NNN) Se (green circles) atoms. The radii of the circles reflect the magnitude of the spectral weight of the particular orbitals to the bands.}
	\label{figure:Figure4}
\end{figure*}

Turning the SOC, lifting the spin degeneracy is expected in the defect states since the inversion symmetry of the ML PtSe$_{2}$ is broken by the formation of the Se-VLD. However, as shown in Figs. 3(a)-3(b), we find that the spin splitting in the defect states is highly anisotropic. Due to the 1D confined states along the extended vacancy line [Figs. 3(c)-(e)], the significant spin splitting is enforced to occur in the defect states along the $\Gamma-Y$ direction [Fig. 3(b)]. However, an extremely small (nearly zero) spin splitting maintains in the defect states along the $\Gamma-X$ direction [Fig. 3(a)], and being completely degenerated when the supercell size perpendicular to the extended vacancy line is increased [see Figs. S1 and S2 in Supplemental Material in detail \cite{Supplementary}]. Since the spin splitting is preserved only along the $\Gamma-Y$ line ($k_{y}$ direction), an ideal 1D SOC is achieved, which is expected to induce a highly spin coherency as recently predicted on the 1D topological defect induced by screw dislocation in semiconductors \cite{LHu}, and hence promising for spintronic applications. Later on, we will show that such typical SOC is reflected by our SOC Hamiltonian derived from the $\vec{k}\cdot\vec{p}$ perturbation theory and symmetry analysis. 

Although the spin splitting in the defect states is subjected to the structural symmetry and the 1D nature of the Se-VLD engineered ML PtSe$_{2}$, it is expected that orbital hybridizations in the defect states play an important role. Here, coupling between atomic orbitals contributes to the nonzero SOC matrix element through the relation $\zeta_{l}\left\langle \vec{L}\cdot \vec{S}\right\rangle_{u,v}$, where $\zeta_{l}$ is angular momentum resolved atomic SOC strength with $l=(s,p,d)$, $\vec{L}$ and $\vec{S}$ are the orbital angular momentum and Pauli spin operators, and $(u,v)$ is the atomic orbitals. Therefore, only the orbitals with the non-zero magnetic quantum number ($m_{l}\neq 0$) will contribute to the spin splitting. By calculating the orbital-resolved of the electronic band structures projected to the atom around the VLD site, we find that large spin splitting in the defect states is mostly originated from the contribution of the $p_{x}+p_{y}$ ($m_{l}=\pm 1$) orbital of the next nearest neighbor (NNN) and nearest neighbor (NN) Se atoms and $d_{x^{2}-y^{2}}+d_{xy}$ ($m_{l}=\pm 2$) orbital of the Pt atom [Fig. 4]. This result is consistent with previous calculations \cite{Li, AbsorA} that introducing a chalcogen vacancy in the ML TMDCs leads to the significant in-plane $p-d$ orbitals coupling, which plays an important role in inducing the large spin splitting in the defect states. 

To further demonstrate the nature of the observed spin splitting in the defect states of the Se-VLD engineered ML PtSe$_{2}$, we show the calculated spin textures around the $\Gamma$ point in Fig. 5(a) for the occupied defect state (DS-1), and in Figs. 5(b)-5(c) for the unoccupied defect states (DS-2, DS-3). It is clearly seen that the spin textures for the upper and lower branch of the spin-split defect states (DS-1, DS-2, DS-3) exhibit a uniform spin configuration along the $x$-direction except for $k_{y}=0$, indicating that the spin polarization is dominated by $\sigma_{x}$ component of spins. However, the spin orientation is fully reversed when $k_{y}$ component of the wave vector is changed to  $-k_{y}$, due to time-reversal symmetry. We reveal that there is no deviation of the spin orientation even at the large $k_{y}$, indicating that the spin textures are highly unidirectional. This peculiar pattern of the spin textures gives rise to the so-called persistent spin textures (PST)\cite{Schliemann,Bernevig}, protecting the spin from the decoherence through the Dyakonov-Perel (DP) spin-relaxation mechanism\cite {Dyakonov, Altmann}. Accordingly, an extremely long spin lifetime is expectable, offering a promising platform to realize an efficient spintronics device. Previously, the PST has been widely studied for semiconductor quantum well (QW) heterostructures \cite{Walser, Schonhuber, Sasaki, Ishihara}. However, achieving the PST in these materials requires a stringent condition for fine-tuning the equal the Rashba and Dresselhaus (RD) SOC parameters\cite{Schliemann, Bernevig}, which is practically non-trivial. On the other hand, only few materials, which is predicted to maintain the PST intrinsicaaly, as recently reported on bulk BiInO$_{3}$ \cite{LLTao}, CsBiNb$_{2}$O$_{7}$\cite{Autieri}, ZnO (10-10) surface\cite{Absor5}, and newly reported 2D materials including ML WO$_{2}$Cl$_{2}$\cite{Ai} and ML group-IV monochalcogenide \cite{Absor3,Absor4}. Therefore, our finding of the PST in the Se-VLD engineered ML PtSe$_{2}$ may provide a distinct advantage over the widely studied PST materials, albeit available intrinsically in a single material.

\begin{figure*}
	\centering
		\includegraphics[width=1.0\textwidth]{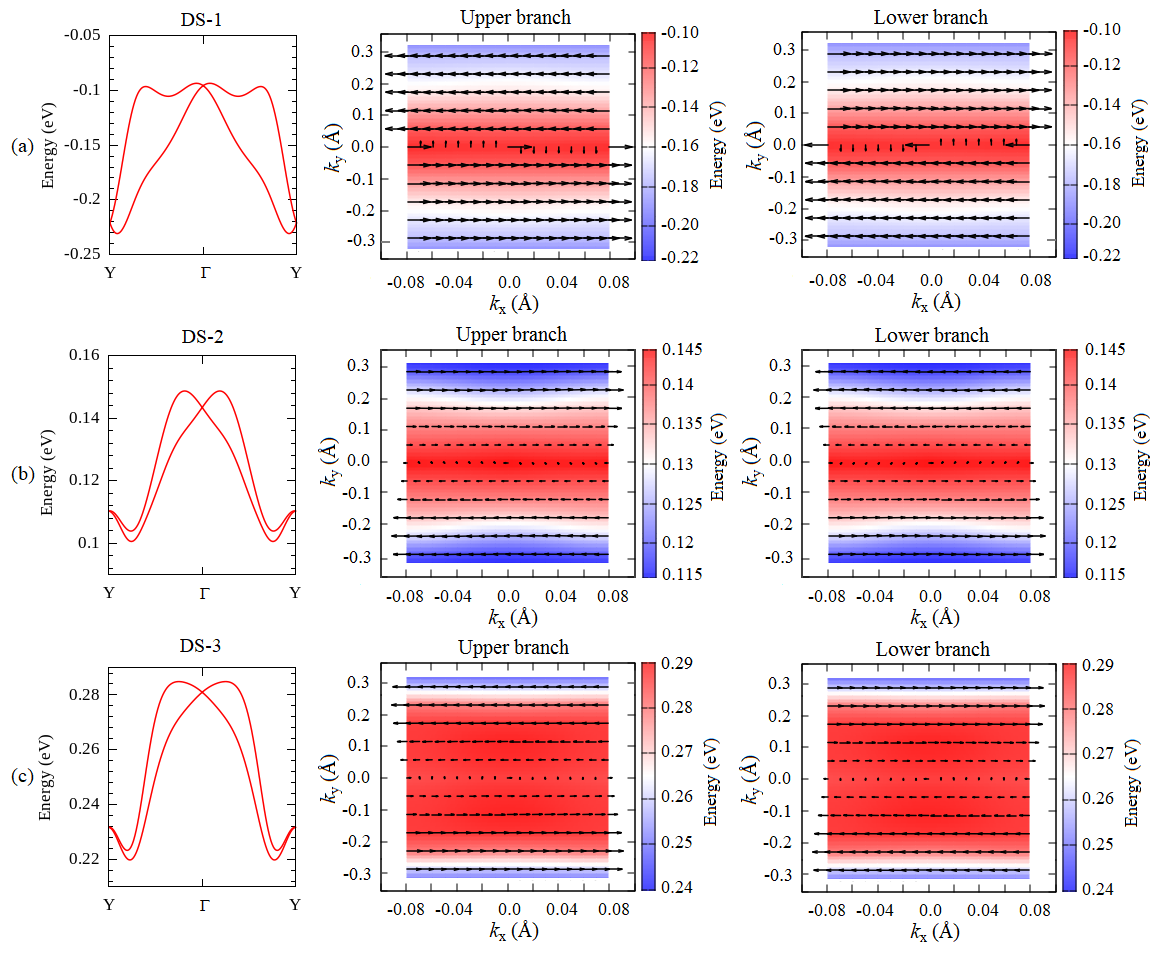}
	\caption{Spin-split bands of the defect states along the $Y-\Gamma-Y$ line corresponding to the spin textures in the momentum space for (a) the DS-1, (b) DS-2, and (c) DS-3. The colour scales represent energy of the band near the defect states, while the black arrows indicate the orientation of the spin polarization in the momentum $k$-space. }
	\label{figure:Figure5}
\end{figure*}

To understand the origin of the observed spin splitting and PST in the defect states of the Se-VLD engineered ML PtSe$_{2}$, we derive a minimal SOC Hamiltonian, $\hat{H}_{\textbf{SOC}}$, by using $\vec{k}\cdot\vec{p}$ perturbation theory combined with symmetry analysis. Although the inversion symmetry is absence in the Se-VLD, it's time-reversal symmetry is conserved at the high symmetry point in the FBZ such as the $\Gamma$ (0,0,0) and Y (0,$\pm$ 0.5, 0) points, leading to the fact that the spin degeneracy remains. When the SOC is introduced, the spin degeneracy is lifted at the $\vec{k}$ away from the time-reversal-invariant points, in which $\hat{H}_{\textbf{SOC}}$ can be derived by $\vec{k}\cdot\vec{p}$ theory. Following Vajna. $et$. $al$., it is possible to construct $\hat{H}_{\textbf{SOC}}$ from the following invariance formulation \citep{Vajna}: 
\begin{equation}
\label{4}
H_{\textbf{SOC}}(\vec{k})= \alpha\left(g\vec{k}\right)\cdot \left(\det(g)g\vec{\sigma} \right),
\end{equation}
where $\vec{k}$ and $\vec{\sigma}$ are the electron's wavevector and spin vector, respectively, and $\alpha\left(g\vec{k}\right)=\det(g)g\alpha\left(\vec{k}\right)$, where $g$ is the element of the point group characterizing the small group wave vector $G_{Q}$ of the high symmetry point $Q$ in the FBZ. 
In Eq. (\ref{4}), we have implicitly assumed that all of the orbital characters at the $Q$ point is invariant under the symmetry transformation in the $G_{Q}$. Therefore, the spin vector, $\vec{\sigma}$, can be considered as a pseudovector. By transforming $\vec{k}$ and $\vec{\sigma}$ as polar and axial vectors, respectively, and sorting out the components of these vectors according to irreducible representation (IR) of $G_{\vec{Q}}$, we can decompose again their direct product into the IR. According to the Eq. (\ref{4}), only the total symmetric IR from this decomposition contributes to $\hat{H}_{\textbf{SOC}}$. Therefore, by using the corresponding tables of the point group, one can easily construct the possible term of $\hat{H}_{\textbf{SOC}}$. 

\begin{table}[ht!]
\caption{Character tabel for $C_{s}$ point group with two elements of the symmetry operation, namely identity operation $E:(x,y,z)\rightarrow(x,y,z)$ and mirror symmetry operation $M_{yz}: (x,y,z)\rightarrow(-x,y,z))$. Two one-dimensional irreducible representations (IR) are shown.} 
\centering 
\begin{tabular}{c c c c c c c c c c c c c c c c c c c c c c c c c}
\hline\hline 
IRs &  & & & &  &  &  & $E$ &  &  &  &  &  &  &  & $M_{yz}$   &   & &  &  &  &   &    &   Linear, Rotation \\ 
\hline 
$A'$ &  &  &  &  &  &  &  & 1 &  &  &  &   &  &  &  & 1   &  &  &  &  &   &   &    &   $y$, $z$, $R_{x}$\\
$A"$ &  & & & &  &  &  & 1 &  & & & &  &  &  & -1   & & & &  &   &   &    &  $x$, $R_{y}$,$ R_{z}$\\
\hline\hline 
\end{tabular}
\label{table:Table 2} 
\end{table}

\begin{table}[ht!]
\caption{Direct product table of the irreducible representations for $C_{s}$ point group.} 
\centering 
\begin{tabular}{c c c c c c c c c c c c c c c c c c c } 
\hline\hline 
 &&&& & & & & &$A'$ &&&& & & & & & $A"$   \\ 
\hline 
$A'$ &&&&& & & & & $A'$ & & & & &&&&& $A"$   \\
$A"$ &&&& & & & & &$A"$ & & & & &&&&& $A'$  \\ 
\hline\hline 
\end{tabular}
\label{table:Table 3} 
\end{table}

In the Se-VLD engineered ML PtSe$_{2}$, the structural symmetry is characterized by only the mirror symmetry operation $M_{yz}: (x,y,z)\rightarrow(-x,y,z)$, and hence belongs to $C_{s}$ point group. Therefore, the small group of the wave vector at the $\Gamma$ point also belongs to $C_{s}$ point group. Due to the 1D nature of the wave function in the defect states [Figs. 3(c)-3(e)], the orbital characters of the wave function is imposed to be invariant under all symmetry operations in the $C_{s}$ point group. Accordingly, $\vec{\sigma}$ can be viewed as a pseudo-vector, which allows us to apply the Eq. (\ref{4}) for deriving $\hat{H}_{\textbf{SOC}}$. On the basis of the character table of $C_{s}$ point group [see Table II], we short out the component of $\vec{k}$ and $\vec{\sigma}$ according to the IR as $A'$: $k_{y}$, $k_{z}$, $\sigma_{x}$ and $A"$: $k_{x}$, $\sigma_{y}$, $\sigma_{z}$. Moreover, from the corresponding table of direct products [Table III], we obtain the third order terms of $\vec{k}$ as $A'$: $k^{3}_{y}$, $k^{3}_{z}$, $k^{2}_{y}k_{z}$, $k^{2}_{x}k_{z}$, $k^{2}_{z}k_{y}$  and $A"$: $k^{3}_{x}$, $k^{2}_{y}k_{x}$, $k_{y}k_{x}k_{z}$, $k^{2}_{z}k_{x}$. However, due to the 1D nature of the defect, all the terms containing $k_{x}$ and $k_{z}$ should vanish. Therefore, according to the table of direct products, the combination of the first order of $\vec{k}$ as well as the third order of $k^{3}$ with $\vec{\sigma}$ that  belongs to $A'$ IR are $k_{y}\sigma_{x}$ and $k^{3}_{y}\sigma_{x}$. This combinations can be generalized for the higher odd $n$th-order of $\vec{k}$, where the only nonzero term is $k^{n}_{y}\sigma_{x}$. By collecting all these terms, we obtain $\hat{H}_{\textbf{SOC}}$ near the $\Gamma$ point as   
\begin{equation}
\label{5}
H_{\textbf{SOC}}(\vec{k})= \alpha_{1}k_{y}\sigma_{x}+\alpha_{3}k^{3}_{y}\sigma_{x} +\alpha_{5}k^{5}_{y}\sigma_{x}+...+\alpha_{n}k^{n}_{y}\sigma_{x},
\end{equation}
where $\alpha_{n}$ is the odd $n$th-order SOC parameter.

\begin{table}[ht!]
\caption{Linear term parameter of the SOC Hamiltonian  $\alpha_{1}$ (in eV\AA) and the wavelength of the PSH $\lambda$ (in nm) calculated on the defect states (DS-1, DS-2, DS-3) of the Se-VLD engineered ML PtSe$_{2}$ compared with those observed on various PSH materials.} 
\centering 
\begin{tabular}{c c c c} 
\hline\hline 
  Systems & $\alpha_{1}$ (eV\AA)  & $\lambda_{PST}$ (nm)  & Reference \\ 
\hline 
Se-VLD in ML PtSe$_{2}$ &   &   &    \\
 DS-1 & 1.14     &   6.33     &   This work \\
 DS-2 &  0.20    &    29.47    &   This work \\
 DS-3 &  0.28   &    28.19    &   This work \\
Interface    &    &  &   \\ 
GaAs/AlGaAs & (3.5-4.9)$\times 10^{-3}$  & (7.3-10) $\times10^{3}$& Ref.\cite{Walser} \\
               & 2.77 $\times10^{-3}$ & 5.5$\times10^{3}$ & Ref.\cite{Schonhuber} \\
InAlAs/InGaAs & 1.0 $\times10^{-3}$&      & Ref.\cite{Ishihara}\\
              & 2.0 $\times10^{-3}$&      & Ref.\cite{Sasaki}\\
LaAlO$_{3}$/SrTiO$_{3}$ & 7.49 $\times10^{-3}$ & 9.8  & Ref.\cite{Yamaguchi}\\
Surface    &    &  &   \\               
ZnO(10-10) surface & 34.78 $\times10^{-3}$& 1.9$\times10^{2}$ & Ref.\cite{Absor5}\\
Bulk    &    &  &   \\ 
BiInO$_{3}$ &1.91 & 2.0 & Ref.\cite{LLTao}\\
CsBiNb$_{2}$O$_{7}$ &0.012-0.014 &  & Ref.\cite{Autieri}\\
2D ML systems   &    &  &   \\ 
$MX$ ($M$:Sn,Ge; $X$:S,Se,Te)  & 0.09-1.7& 1.82-1.5$\times10^{2}$  & Ref.\cite{Absor4}  \\
Halogen-doped SnSe & 1.6-1.76 & 1.2-1.41 & Ref.\cite{Absor3}\\
WO$_{2}$Cl$_{2}$ &  0.9     &      & Ref.\cite{Ai}\\
\hline\hline 
\end{tabular}
\label{table:Table 4} 
\end{table}

It is revealed from Eq. (\ref{5}) that $\hat{H}_{\textbf{SOC}}$ is characterized only by the one component  of the wave vector, $k_{y}$, and the spin vector, $\sigma_{x}$. Therefore, $\hat{H}_{\textbf{SOC}}$ preserves an ideal 1D SOC, as recently predicted on the 1D topological defect induced by screw dislocation in semiconductors \cite{LHu}. Accordingly, the spin splitting is expected to occur only along the $k_{y}$ direction ($\Gamma-Y$ line), which is in agreement with the band dispersion in Fig. 3(b) obtained from our DFT calculations. On the other hand, it is clearly seen from Eq. (\ref{5}) that the only non-zero spin component in the $\hat{H}_{\textbf{SOC}}$ is $\sigma_{x}$, indicating that the spin orientation is collinear to the $x$-axis, and thus maintaining the PST. Importantly, the 1D nature of the SOC found in the $\hat{H}_{\textbf{SOC}}$ is conserved up to the higher-order term of $k_{y}$, indicating that the formation of the PST remains at the larger $k_{y}$ without any deviation of the spin polarization, which is consistent well with the highly unidirectional spin textures shown in Figs. 5(a)-5(c).

We highlight the main difference of our derived $\hat{H}_{\textbf{SOC}}$ given in Eq. (\ref{5}) from the widely studied 2D Rashba-Dresselhaus (RD) SOC effect supporting the PST. In the 2D RD-SOC, the SOC Hamiltonian can be expressed as $\hat{H}_{\textbf{SOC}}=\vec{\Omega}\cdot\vec{\sigma}$, where $\vec{\Omega}$ is the spin-orbit field (SOF) defined as $\vec{\Omega}=(\alpha_{R} k_{y}+\alpha_{D} k_{x}, -\alpha_{R} k_{x}-\alpha_{D} k_{y},0)$, with $\alpha_{R}$ and $\alpha_{D}$ are the Rashba and Dresselhaus  SOC parameters, respectively. Such SOC leads to a chiral spin texture characterized by the spin-momentum locking property, which is known to induce the undesired effect of causing spin decoherence\cite {Dyakonov}. The PST occur when the SOF $\vec{\Omega}$ is unidirectional, which can be achieved, in particular, if the magnitude of $\alpha_{R}$ and $\alpha_{D}$ are equal \cite{Schliemann, Bernevig}, as previously observed in conventional semiconductor quantum wells (QWs)\cite{Walser, Schonhuber, Sasaki, Ishihara}. However, the stringent condition of $\alpha_{R}=\alpha_{D}$ is difficult to satisfy since it requires matched the QW width, doping level, and application of an external gate bias. On the other hand, the formation of the PST driven by the 2D RD-SOC is generally broken by higher-order term of $k$ \cite{Walser, Ishihara}, which has a considerable effect of reducing the spin coherency. In contrast, for our derived $\hat{H}_{\textbf{SOC}}$ in Eq. (\ref{5}), the 1D nature of the SOC enforces the SOF being always unidirectional in the $x$-direction whatever the order of $k_{y}$ is. Therefore, the PST maintains even at the larger $k_{y}$, giving advantages to the significantly higher degree of spin coherency than that of the 2D RD-SOC. 

For a quantitative analysis of the predicted PST in the Se-VLD engineered ML PtSe$_{2}$, we calculate the SOC parameter $\alpha_{1}$ associated with the linear term of $\hat{H}_{\textbf{SOC}}$ given in Eq. (\ref{5}), and compare the result with a few selected PSH materials. By fitting the DFT bands of the defect states along the $\Gamma-Y$ line, we find that the calculated $\alpha_{1}$ for the DS-1 state is 1.14 eV\AA\, which is larger than that for the DS-2 state (0.20 eV\AA) and DS-3 state (0.28 eV\AA) [see Table IV]. However, this value is much larger than that of the semiconductor QWs systems such as GaAs/AlGaAs \cite{Walser,Schonhuber} and InAlAs/InGaAs \cite{Sasaki,Ishihara}, ZnO (10-10) surface\cite{Absor5}, and strained LaAlO3/SrTiO3 (001) interface \cite{Yamaguchi}. Even, this value is comparable with that of the bulk CsBiNb$_{2}$O$_{7}$\cite{Autieri}, BiInO$_{3}$ \cite{LLTao}, and newly reported 2D materials including the ML WO$_{2}$Cl$_{2}$\cite{Ai}, and ML group-IV monochalcogenide \cite{Absor3,Absor4} [see Table IV]. Remarkably, the associated SOC parameters found in the defect states of the Se-VLD engineered ML PtSe$_{2}$ are sufficient to support the room temperature spintronics functionality.  

The observed PST in our defective system may result in a spatially periodic mode of the spin polarization emerging in the crystal known as a persistent spin helix (PSH) \cite{Bernevig}. The corresponding spin-wave mode is characterized by the wavelength of $\lambda=(\pi\hbar^{2})/(m_{\Gamma-Y}^{*}\alpha_{1})$\cite{Bernevig}, where $m_{\Gamma-Y}^{*}$ is the carrier effective mass along the $\Gamma-Y$ direction. By fitting the band dispersion in the defect states along the $\Gamma-Y$ line, we find that the calculated $m_{\Gamma-Y}^{*}$ is -0.21$m_{0}$ for the DS-1 state, while it is found to be 0.25$m_{0}$ and 0.19$m_{0}$ for the DS-2 and DS-3 states, respectively, where $m_{0}$ is the free electron mass. The negative (positive) value of $m_{\Gamma-Y}^{*}$ characterizes the effective mass of the hole (electron) carriers in the occupied (unoccupied) defect states. The resulting wavelength $\lambda$ is 6.33 nm for the DS-1 state, which is one order smaller than that for the DS-2 (29.47 nm) and DS-3 (28.12 nm) states [see Table IV]. Specifically, the calculated  $\lambda$ for DS-1 state is comparable with that reported on the bulk BiInO$_{3}$ \cite{LLTao} and ML group IV monochalcogenide \cite{Absor3,Absor4} [see Table IV], rendering that the present system is promising for nanoscale spintronics devices.

Thus far, we have found that the PST is achieved in the defect states of the Se-VLD engineered ML PtSe$_{2}$. In particular, the PST with the largest strength of the spin splitting  ($\alpha_{1}=1.14$ eV\AA) is observed in the DS-1 state, indicating that the PSH will be formed when the hole carriers are optically injected into the occupied defect state of the Se-VLD engineered ML PtSe$_{2}$. Since the wavelength of the PSH in the DS-1 state ($\lambda=6.33$ nm) is substantially small, it is possible to resolve the features down to the tens-nm scale with sub-ns time resolution by using near-filled scanning Kerr microscopy\cite{Rudge}. In addition, due to the sizeable spin splitting in the DS-1 state, the two states with opposite spin orientation at $k_{y} $ and -$k_{y}$ are expected to induce large Berry curvature with opposite sign. By using polarized optical excitation technique, it is possible to create different hole population between these two states. Therefore, a charqe Hall current can be measured similar to the valley Hall effect recently discovered in TMDCs\cite{Mak}. As such, our findings of the large spin splitting in the Se-VLD engineered ML PtSe$_{2}$ maintaining the PST is useful for spintronic applications.

\section{CONCLUSION}

The effect of the line defect on the electronic properties of the ML 1T-TMDCs has been systematically investigated by employing the first-principles DFT calculations. Taking the the ML PtSe$_{2}$ as a representative example, we have considered the most stable form of the vacancy line defects (VLDs), namely the Se-VLD. Our band structures analysis have shown that the midgap defect states are observed in the Se-VLD, exhibiting the dispersive characther of the bands along the $\Gamma-Y$ direction. By taking into account the SOC in the DFT calculations, we have revealed a sizable spin splitting in the defect states of the Se-VLD, which is mainly derived from the strong hybridization between the in-plane $p-d$ orbitals. Importantly, we have observed a highly unidirectional spin configuration in the spin split defect states, giving rise to the so-called persistent spin textures (PST) \cite{Schliemann, Bernevig}, which protects the spin from decoherence and induces an extraordinarily long spin lifetime. Moreover, by using $\vec{k}\cdot\vec{p}$ perturbation theory supplemented with symmetry analysis, we have demonstrated that the emerging of the spin splitting maintaing the PST in the defect states is subjected to the inversion symmetry breaking together with the 1D nature of the Se-VLD engineered ML PtSe$_{2}$. Recently, the defective ML 1T-TMDCs has been extensively studied \cite {AbsorA, WZhang, Zulfiqar, Kuklin}. Our study clarifies that the line defect plays an important role in the spin-splitting properties of the ML 1T-TMDCs, which could be highly important for designing spintronic devices.

We emphasized here that our proposed approach for inducing the large spin splitting by using the line defects is not only limited on the ML PtSe$_{2}$ but also can be extendable to other ML 1T-TMDCs systems such as the ML Pd$X_{2}$ ($X$= S, Se, Te)\cite{Kuklin}, ML Sn$X_{2}$\cite{Gonzalez},  ML Re$X_{2}$\cite {Horzum}, and ML (Zr/Hf)$X_{2}$\cite{CaiC}, where the structural and electronic structure properties are similar. Recently, manipulation of the electronic properties of these particular materials by introducing the defect has been reported\cite {Kuklin}. Therefore, it is expected that our predictions will stimulate further theoretical and experimental efforts in the exploration of the spin-splitting properties of the ML TMDCs, broadening the range of the 2D materials for future spintronic applications.

\begin{acknowledgments}

The first author (M.A.U. Absor) would like to thanks Nanomaterial Research Institute, Kanazawa University, Japan, for providing financial support during his research visit. This work was partly supported by Grants-in-Aid on Scientific Research (Grant No. 16K04875) from the Japan Society for the Promotion of Science (JSPS) and a JSPS Grant-in-Aid for Scientific Research on Innovative Areas ”Discrete Geometric Analysis for Materials Design” (Grant No. 18H04481). This work was also partly supported by PDUPT Research Grant (2020) funded by the Ministry of Research and Technology and Higher Education (RISTEK-DIKTI), Republic of Indonesia. Part of this research was also supported by the MIRA Research Grant (2019-2020) funded by the Ministry of Education and Cultures, Republic of Indonesia. The computation in this research was performed using the supercomputer facilities at RIIT, Kyushu University, Japan.

\end{acknowledgments}

\bibliography{Reference1}


\end{document}